# A Review of Phasor Measurement Unit Requirements and Monitoring Architecture Practices for State Estimation at Distribution Systems

Panayiotis Moutis
*DEPsys SA*
Puidoux, Switzerland
panayiotis.moutis@depsys.ch

Omid Alizadeh-Mousavi
*DEPsys SA*
Puidoux, Switzerland
panayiotis.moutis@depsys.ch

*Abstract*—Distribution system and market operators will need or be required to monitor closely distribution systems, due to the presence of multiple actors, such as generating and storage units, as also active loads. The reason for this is the effect of these latter actors to system operation as also to ensure they follow their commitments in the deregulated market frameworks. This paper discusses why the traditional approach to state estimation as this has been developed for transmission systems cannot be extended to distribution systems. Some preliminary proposals are made, and a high-level architecture is presented as part of an on-going research work on the matter.

*Keywords—Distribution system, distributed generation, phase measurement unit, state estimation, voltage angle.*

## I. Introduction

Multiple market incentives have been developed and offered during the past few years, in order to promote the wide deployment of renewable energy sources [1, 2]. Backed by tariff and other subsidizing mechanisms [3, 4] various types and sizes of Distributed Generation (DG) units [5] have been installed mostly at the Distribution System (DS) levels throughout the world. With the roll-back of these financial support mechanisms [6], DG is expected to be aggregated and enter the energy and ancillary services markets accordingly. Lately, Storage Systems (SS) have also been employed either complementary to DG or as stand-alone projects aiming to gain revenues from services and price patterns noted in the said deregulated markets [7]. Loads have also been enabled to participate in this business through demand response programs [8]. All these points describe a landscape of a multitude of actors with very different characteristics, conflicting interests and operating at a part of the grid that is barely monitored (if not at all), although it serves almost all customers in this business (in the traditional sense), the loads. This landscape implies that DS are exposed to many power quality and reliability issues [9, 10], with no effective oversight that can respond to events, improve customer service and experience, or prevent efficiently any of these issues.

Although pricing mechanisms may be employed by Distribution System and Market Operators (DSMO), in order to drive all the aforementioned actors in certain behaviors that will optimize DS operation [11], the stochastic nature both of loads as also of many DG units based on renewables [12], render such an approach dubious. At the same time, the size of DS implies that any types and numbers of random errors might occur, which cannot possibly be foreseen [5]. DSMO need to adopt a more controlled approach for the operation of DS and a first step towards that direction has been with projects replacing vast numbers of typical meters with smart ones and other DSMO ventures deploying equipment to monitor DS transformers [13]. Nevertheless, these projects focus on a binary view of the system assets under control; i.e. the equipment is either operational or not. Many failures, errors, system disconnects, and power quality matters may be handled and prevented through proper measures, provided there exists the right knowledge of the system status. To this end, the potential of employing State Estimation (SE) at DS has been gaining momentum with studies, publications and research initiatives [13-20]. SE is the algorithmic and hardware framework, which, based on the network topology of a power system (admittance matrix of the lines, transformers, etc.), processes online measurements from various points in that system to calculate an as close to real-time as possible power flow solution/representation [21].

Although numerous papers have been published on SE at DS from the perspectives of the mathematical formulations that can best serve the purpose, and the placement of measuring devices that will minimize infrastructure cost and enhance the performance of the methods initiatives [13-20], only few works take a more applied approach [22, 23]. In those works, the proposed ideas are practically valid, and the gathered results are valuable, while a hardware design for a micro Phase Measurement Unit (uPMU – the core measurement device for any SE application) has been a groundbreaking innovation for SE at DS. However, there are many considerations that require for or would be better handled with architectures and methods, which are not common in the traditional sense of SE (employed in transmission systems). This paper aims to discuss specifically those considerations and make some first suggestions in a direction that will best serve SE for DS.

In Section II the main drivers that justify the practical need for SE in DS are presented. In Section III, drawing from the aforementioned drivers and comparing with practices of the traditional SE at the transmission systems, a few suggestions are made, first, about the preferred characteristics of the PMUs to be deployed for that matter, and, secondly, about the

This work has received funding from the European Union's Horizon 2020 research and innovation programme under the Marie Sklodowska-Curie grant agreement No 797451.

architecture of the SE methodology. Section IV concludes this work and offers future directions of the on-going research on this topic.

## II. Towards State Estimation at Distribution Systems

The academia and the industry have identified, through extensive R&D, various needs, applications and challenges in the operation and quality of service offered at the DS, that concretely justify the need for SE.

Firstly, owing to environmental concerns, Renewable Energy Sources (RES) have been widely promoted and, thus, installed at DS. Despite their generally positive economic and environmental impacts, most RES (especially the prominent ones – wind generation and photovoltaics) are characterized by volatility [24], which causes excessive activation and, thus, wear of infrastructure owned by the network operator (regulators and on-load tap changers of transformers) [25]; not mentioning additional power quality issues [26].

Secondly, although it is agreeable that wide penetration of Distributed Generation (DG) postpones investments in system reinforcement [5], it may not be a guaranteed outcome, while it is also unclear for how long and/or for which parts of a DS. There is a two-fold reason to this. On one hand, wide DG deployment may lead to unacceptable voltage rise (requirement for compensation) [25, 27], and/or upstream/reverse power flows (requirement for bi-directional switching equipment) [28]. On the other hand, certain parts of the DS may still become congested (reactive power served into the DS from the overlying grid, unlike in recently proposed standards [29]) or suffer poor voltage profiles (uneven load increase over time) [30].

Thirdly, employing SE at DS will increase system reliability indicators, since it will improve handling of faults by locating them [21, 31] and responding to them more efficiently [32]. Additionally, factors and parameters that could cause reliability concerns may be closely monitored and, thus, through well-informed DSMO actions, prevent imminent faults.

Fourthly, DS reconfiguration will also be greatly favored thanks to SE at this level [14, 19, 33-35]. The SE will facilitate the DS operator in monitoring the conditions, which dictate the need to reconfigure a system, and assess the effects of every reconfiguration action to the operating profile of the said system. On the same issue, reconfiguration may be realized remotely by ensuring that coupling conditions between connecting parts of a system are correctly met.

Lastly, if SE is active on a given DS, standards that limit the greater penetration of DG (stricter criteria calculated offline and prior to the installation of such units), may be applied in a softer manner. For example, voltage deviations averaged over specific time intervals [9, 36] may be contained through the online control of the DG units that cause them.

Additional arguments to the same direction, such as the effects by the increased use of plug-in electric vehicles, the demand response programs both for residential customers and industrial loads, and others are not analyzed separately, since they can be conveniently categorized under the aforementioned broader concerns. It has been made clear that SE of DS is a current problem of utmost importance, in order to monitor and reduce the effects of RES while further promoting them, to log the behavior of DS with regards to long-term investment/infrastructure decisions, and to enhance end-customer load service and offered quality.

## III. Practical Concerns for State Estimation at Distribution Systems

### A. Expected Requirements for Phase Measurement Units at Distribution Systems

PMUs are at the heart of SE, since they are the preferred (although not exclusively employed) measuring devices on which the whole methodology is based [37]. Their accuracy, their refresh rates, and their location in the monitored power system are the main concerns that have been most thoroughly studied, researched, and standardized [19, 33, 35, 38-40]. It is valuable to first note some of the standardized features of PMUs employed at the transmission systems that cannot be plainly justified for the requirements of SE at DS.

Electric frequency and rate of change of frequency (ROCOF) have been specified to be measured with very small margins of error, which were considered tight and have had to be relaxed from the original version of the same standard series [37]. Although these measurements are crucial to Transmission System Operators (TSO) for various reasons, they are practically unimportant to DSMOs. The latter entities have no stake, obligation or considerable effect in events that may be detected through frequency deviations or their rates. Although recent standards are describing more active roles for DG units in these control domains [36] and previous publications have established such capabilities [12, 41], their behaviors are not expected to be supervised. At the same time, no market frameworks or grid codes describe a setting in which large fleets of DG units can be actively involved and efficiently respond to relieving the system of frequency deviations (the so-called primary frequency control) [42]. As of the above, a credible frequency measurement should be considered a minimum requirement for monitoring the effect of generation-demand imbalances downstream the DS, but ROCOF can be of little insight to DSMOs. An important remark here is the case of the Australian black-out, which was caused, among other reasons by rapid changes in frequency, due to lack of regulating inertia [43]. The ROCOF that was observed (greater than 5Hz/s) would require for every available controlled asset throughout the system to respond in an unprecedented manner, to prevent the event. This extreme case cannot unequivocally justify strict requirements for ROCOF measurements by PMUs at DS.

Another error metric set forward by the same standard is the Total Vector Error (TVE), calculated as follows:

$$\text{TVE} = \sqrt{\frac{(\tilde{v}_r - \hat{v}_r)^2 + (\tilde{v}_i - \hat{v}_i)^2}{\tilde{v}_r^{\,2} + \tilde{v}_i^{\,2}}} \qquad (1)$$

Where values under tildes are the real values of the waveforms measured, while values under hats the measurement outputted by the PMU device. TVE is required to be limited to ±1% (with reference to nominal voltage magnitude values). For the case of DS, TVE could be required to be measured with an even smaller error bound. The reason for this suggestion is that relevant power quality standards at this level of the grid allow for a ±10% voltage deviation (or less) around its nominal value [9, 44]. The required TVE represents 1/20th of this interval and can have ambiguous effects to the corrective measures employed by DSMOs or to the automatic response from transformers equipped with on-load tap changers (if PMUs are integrated to the control topology) [25].

A particularly disruptive proposal with regards to the measurement requirements from a PMU employed at a DS concerns the voltage angles. Although in the traditional SE framework, a PMU is expected to estimate an accurate voltage angle, for DS this might not be a compelling requirement. The vast majority of DS are operated radially [30], hence the voltage magnitude (compared between/among buses) is an adequate measure of whether power is persistently injected or absorbed at any given point or part of a DS. The term persistently is used in light of the extent to which a DS is monitored through PMUs; unless adequate and properly selected (covering 'equally' the whole DS) control points are measured, parts of the DS will be estimated poorly. The uPMUs, developed specifically for SE in DS, are characterized by particularly high angle measurement precision, albeit at a very high capital cost for every uPMU [22]. At the same time, Optimal Power Flow methods (which are, in mathematical context, the basis of any SE method) tailored for DS have been boasting limited errors in voltage magnitude calculations, while not concerned with voltage angle consideration [45]. System reconfiguration is the only functionality requiring good voltage angle estimate [23], to realize all switching actions. Nevertheless, the switching operations in DS reconfigurations are not arbitrarily employed at any node, but only where such switches have been previously installed. In this sense, voltage angle monitoring in DS is not a strict and general prerequisite for PMUs employed at this level of the system and can be either ignored or specified in case-by-case approach.

### B. State Estimation Architecture and Realization for Distribution Systems

Any TSO assesses SE data gathered and organized based on SCADA systems in an online and high refresh rate set-up [46]. Transient phenomena, inter-area exchanges and other events that are predominantly dynamic in their nature need to be monitored constantly, anticipated, and handled before they cause instability or market irregularities. In contrast to this, most of the phenomena expected to occur and assessed by DSMOs are not dynamic in nature. Voltage deviations from nominal values, disconnects of part(s) of a DS, and activation of DSMO equipment due to DG volatility, to mention a few, are affecting a DS in the time scales of minutes and are also expected to be handled in similar time frames [9]. To this end, high refresh and reporting rates are an excessive requirement that can be moderated, accordingly. It is interesting to note that the previously mentioned R&D effort in SE for DS has highlighted the high telecommunications costs incurred by daily loads of 0.5 GB of transmitted information over cellular networks for realizing an online SE approach [22].

Further to this last point, the central SCADA SE architecture needs to be rethought. Any moderately sized DSMO may be responsible for dozens, if not hundreds, of DS spread through geographically vast areas and spanning, at least, a few tens of buses each. The sheer amount of information required to be handled by any central topology, as also the concerns of single-point failure lead to the consideration of alternative approaches, prominent among which are the distributed methods as also data compression techniques [47]. Such methods of SE will only require limited exchange of data between neighboring PMUs, which might be able to be realized over communication protocols/channels other than cellular networks [48]. Moreover, since most of the challenges for SE at DS, as described in the previous section, do not require for constant online monitoring, the methods may be informed for power quality, stability and reliability standards and alert the DSMO only in such cases, thus further reducing communication costs.

### IV. CONCLUSIONS AND FUTURE WORK

It has been made clear that SE at DS cannot be assessed in the perspective of SE as this has been traditionally employed at transmission systems. The here described technical applications that drive the need for SE at DS, specify some high-level characteristics both for the PMUs and the SE architecture and realization. Nevertheless, some of the discussed features can be described as disruptive and debatable, while their effects and impact can only be assessed by the major stakeholders in the field which are the DSMOs.

To this end, this R&D effort will be immediately addressing DSMOs with an extensive survey. The survey questionnaires have been organized in three parts. One part is focused on whether and to what extend any measurement infrastructure (either PMU, RTU, energy meters, etc.) is installed and operating within the responsibility of the surveyed DSMO. The second part of the said questionnaire focuses on a projected assessment of employing and operating extensive SE at any part of the DS of a DSMO. The last part of the survey aims to identify incentives, policies and initiatives that may promote SE at DS and for what reasons. The results of this survey will be jointly reviewed with the observations and remarks gathered in this work to formulate a more definitive approach to the credible requirements and expectations for SE at DS.